\documentstyle[epsfig,sprocl]{article}

\def\Journal#1#2#3#4{{#1} {\bf #2}, #3 (#4)}

\def\AP{\em Ann. Phys.}

\def\NPA{{\em Nucl. Phys.} A}

\def\PLB{{\em Phys. Lett.}  B}
\def\PR{\em Phys. Rev.}
\def\PRL{\em Phys. Rev. Lett.}
\def\PRC{{\em Phys. Rev.} C}

\def\ZP{\em Z. Phys.}

\def\be{\begin{equation}}
\def\ee{\end{equation}}
\def\bea{\begin{eqnarray}}
\def\eea{\end{eqnarray}}

\begin{document}

\title{VECTOR MESON
PROPAGATION IN DENSE MATTER~\footnote{This talk is dedicated to
Mannque Rho on the occasion of his 60th birthday.}}

\author{Bengt Friman}

\address{Gesellschaft f\"ur Schwerionenforschung (GSI)\\
D-64220 Darmstadt, Germany\\
\&\\Institut f\"ur Kernphysik, TU Darmstadt\\
D-64289 Darmstadt, Germany\\E-mail: b.friman@gsi.de}

\maketitle\abstracts{The properties of vector mesons in nuclear matter
are discussed. Results for the momentum dependence of the
$\rho$-meson self energy in matter due to $\rho N$ interactions are
presented, and consequences for the production rate of $e^+e^-$
pairs in hot and dense matter are discussed. I also examine the
constraints imposed by elementary processes on the widths of $\rho$
and $\omega$ mesons in nuclear matter, and conclude that in recent
models the in-medium widths of these mesons are overestimated.}

\section{Introduction}
In 1991 Gerry Brown and Mannque Rho proposed a universal scaling
law for effective hadron masses in matter, which connects the
hadron masses with the quark condensate, an order parameter for the
spontaneous breaking of chiral symmetry~\cite{BR}. According to the
scaling law, often referred to as Brown-Rho or BR scaling, the
in-medium hadron masses, except for those of the pseudo-scalar
mesons, are reduced when the quark condensate is reduced, i.e.,
when the chiral symmetry is restored. Thus, if BR scaling is
correct, it opens the possibility to experimentally explore the
restoration of chiral symmetry in dense and hot matter, provided
one can identify experimental signatures for the modification of
the effective masses of hadrons. This idea has triggered a lively
discussion on the properties of hadrons in matter.

The electromagnetic decay of the vector mesons into $e^+e^-$ and
$\mu^+\mu^-$ pairs makes them particularly well suited for
exploring the conditions in dense and hot matter in nuclear
collisions. The lepton pairs provide virtually undistorted
information on the mass distribution of the vector mesons in the
medium. Because of this, the discussion has concentrated on the
in-medium properties of the $\rho$ and $\omega$ mesons and the
corresponding signatures in the lepton-pair spectrum.

The universal scaling conjecture is supported by the QCD sum rule
calculations of Hatsuda and Lee \cite{Hatsuda-Lee}, who find a
strong reduction of the $\rho$- and $\omega$-meson masses in
nuclear matter. However, recently the results of Hatsuda and Lee
were questioned in the work of Klingl, Kaiser and Weise \cite{KKW}.
Klingl {\em et al.} find that the vector meson widths, which were
ignored in the work of Hatsuda and Lee, play an important role in
the QCD sum rules and that the QCD sum rule for the $\rho$ meson is
satisfied by a model, where its width is strongly enhanced in
nuclear matter, while its energy remains almost unchanged. For the
$\omega$ meson they find a strong downward shift of the energy and
a relatively large enhancement of its width. The tradeoff between
the vector-meson energy and width in the QCD sum rules was then
explored in more detail by Leupold, Peters and Mosel \cite{LPM}.

The lepton-pair spectrum in nucleus-nucleus collisions at SPS
energies exhibits a low-mass enhancement compared to proton-proton
and proton-nucleus collisions \cite{CERES,Helios}. A quantitative
interpretation of the lepton-pair data can be obtained within a
scenario, where the effective vector-meson masses are reduced in a
hadronic environment \cite{likobr,cassing}. On the other hand,
attempts to interpret the low-mass enhancement of lepton pairs in
terms of many-body effects also yield good agreement with the data
\cite{CRW,CBRW}. In these calculations the broadening of the $\rho$
meson in nuclear matter due to the interactions of its pion cloud
with the medium \cite{HFN,CS,AK,KKW} and the momentum dependence of
the $\rho$-meson self energy due to the coupling with
baryon-resonance--nucleon-hole states \cite{FP,PLPM} are taken into
account. At least superficially, this mechanism seems to be
completely decoupled from the BR-scaling scenario.

In this talk I discuss the many-body effects in the light of
recent theoretical developments. I also examine the constraints on
the imaginary part of the vector meson self energy in nuclear
matter that can be derived from elementary reactions, like $\pi
N\rightarrow\rho/\omega N$.

\begin{figure}[t]
\center{\epsfig{file=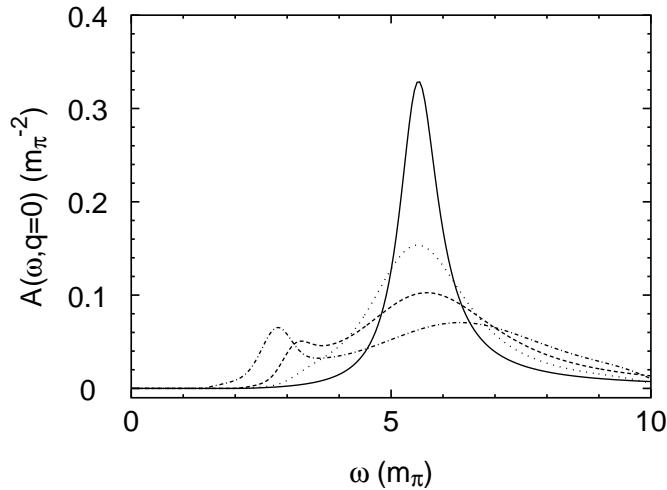,width=72mm,angle=90}}
\caption{\label{fig1} The spectral function of
a $\rho$ meson at rest in nuclear matter at density $\rho= 0$
(full), $\rho_0$ (dotted), $2 \rho_0$ (dashed) and $3 \rho_0$
(dash-dotted line) (from ref.~11).}
\end{figure}

\section{The pion cloud}

The $\rho$ and $\omega$ mesons couple strongly to two and three
pion states, respectively~\cite{PDG}. Consequently, both are
surrounded by a pion cloud. Since pions interact strongly with
nucleons, the pion clouds of the vector mesons are modified in
nuclear matter. For a $\rho$ meson at rest in nuclear matter this
effect was explored several years ago by Herrmann {\em et al.}
\cite{HFN}, Chanfray and Schuck \cite{CS} and by Asakawa {\em et
al.}~\cite{AK}. In these calculations conservation of the isospin
current, which acts as a source for the $\rho$-meson field, was a
guiding principle. The main effect considered was the dressing of
the pion by the excitation of $\Delta (1232)$-hole states. In spite
of differences in the details of the models, all groups came to
similar conclusions, namely that the $\rho$-meson width is enhanced
in matter, but that its energy does not change appreciably.
Furthermore, at high densities ($\rho_N = 2-3\rho_0$), a new branch
appears at an energy of roughly 450 MeV. This branch corresponds to
the decay of the $\rho$ meson into a pion and a $\Delta$-hole
state.

\begin{figure}[t]
\center{\epsfig{file=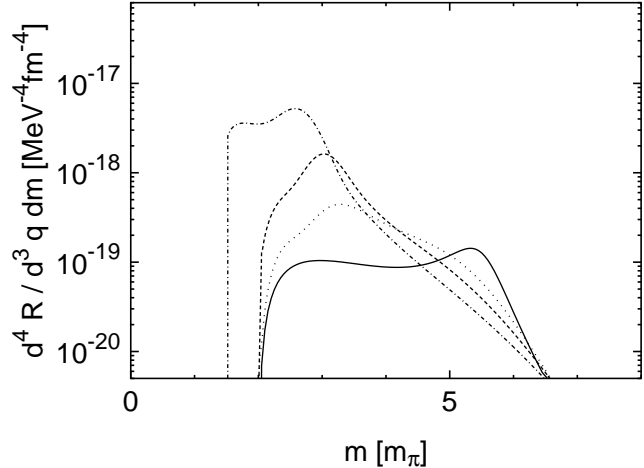,width=77mm,angle=90}}
\caption{\label{fig2} The production rate for lepton pairs in
nuclear matter at T = 100 MeV corresponding to the spectral
function shown in Fig.~\ref{fig1}. The densities are $\rho= 0$
(full), $\rho_0$ (dotted), $2\rho_0$ (dashed) and $3
\rho_0$ (dash-dotted line)(from ref.~11).}
\end{figure}

In Fig.~\ref{fig1} I show the resulting $\rho$-meson spectral
function of ref.~\cite{HFN} and in Fig.~2 the corresponding lepton
pair production rate from $\pi^+\pi^-$ annihilation in hadronic
matter at a temperature of T = 100 MeV, and different densities.
Note that the broadening of the $\rho$ and the new branch leads to
an enhancement of the production rate for low-mass lepton pairs.

Recently Klingl {\em et al.}~\cite{KKW} constructed a
vector-meson--baryon effective interaction of finite range based on a
chirally symmetric Lagrangian. In this model the vector-meson--nucleon
interaction is to a large extent mediated by the meson cloud of the
vector mesons, much like in the earlier work~\cite{HFN,CS,AK}. The
resulting in-medium spectral functions satisfy the QCD sum rules. In
agreement with the earlier calculations, Klingl {\em et al.}  find
that a $\rho$ meson at rest in nuclear matter is broadened, but that
its energy remains almost unchanged. On the other hand, the energy of
the $\omega$ meson is shifted down appreciably and its width is
enhanced substantially. Finally, for the $\phi$ meson they find almost
no change of its energy and a moderate enhancement of its width. In
this work the vector-meson--nucleon scattering amplitude is
renormalized in the limit corresponding to Compton scattering of
long-wavelength photons. Thus, the model involves an extrapolation
from the photon point $q^2=0$ to $q^2=m_V^2$. As we shall see below,
this extrapolation, which is governed by the form factors of the
model, turns out to be hazardous.

\begin{figure}[t]
\center{\epsfig{file=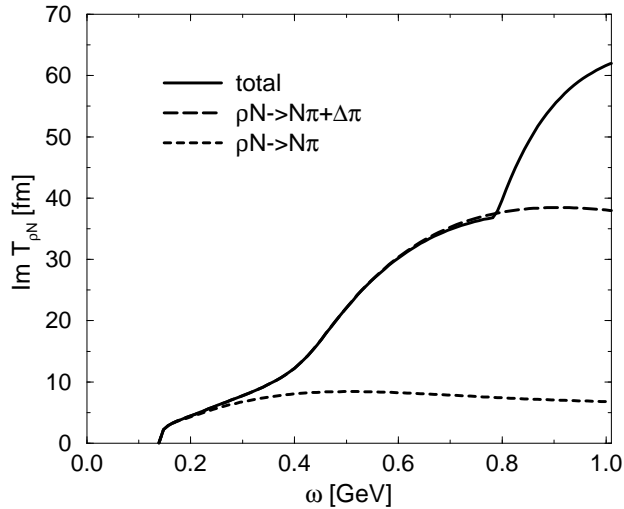,height=75mm}}
\caption{\label{Irho} The imaginary part of the $\rho$-nucleon
T-matrix in the model of Klingl {\em et al.} (ref.~3).}
\end{figure}

The imaginary parts of the $\rho$-nucleon and $\omega$-nucleon
T-matrices obtained by Klingl {\em et al.}~\cite{KKW} are shown
in Fig.~3 and 4. They define a complex scattering length through
\begin{equation}
\label{scatt-length}
a_{VN}=\frac{M_N}{4\pi(M_N+m_V)} T_{VN}(\omega=m_V).
\end{equation}
The imaginary parts of the scattering lengths are related to the
existence of inelastic channels, the $\pi N$ and $\pi\pi N$
channels, which are open at the vector-meson--nucleon threshold.
The complex $\rho N$ and $\omega N$ scattering lengths quoted by
Klingl {\em et al.} are $a_{\rho N\rightarrow\rho N}=(0.04 + i\,
1.62)$~fm and $a_{\omega N\rightarrow\omega N}=(3.34 + i\,
2.1)$~fm, respectively. As discussed in section 4, one can relate
the scattering lengths to the vector-meson self energies in nuclear
matter at low density. Thus, the imaginary parts of the scattering
lengths correspond to very large in-medium enhancements of the
$\rho$ and $\omega$ widths. The values of the scattering lengths
imply that on the free mass shell $\Delta\Gamma_\rho \simeq 300$
MeV and $\Delta\Gamma_\omega \simeq 390$ MeV at normal nuclear
matter density. The real part of the $\omega N$ scattering length
indicates that the $\omega$-meson potential in nuclear matter is
strongly attractive, while the small real part of the $\rho N$
scattering length corresponds to an almost vanishing shift of the
$\rho$-meson energy in a nuclear medium. If the energy of the
$\omega$ meson is reduced in matter due to the attractive
potential, its in-medium width is of course smaller than the value
quoted above. Both the real and imaginary parts of the $\phi N$
scattering length are small in this model.

Using Eq.~(\ref{scatt-length}) and the results shown in Fig.~3 and 4
one can extract the contribution from the $\pi N$ channel to the
imaginary parts of the scattering lengths. I find $\mbox{Im}\, a_{\rho
N\rightarrow\rho N}^{(\pi N)}\simeq 0.35$ fm and $\mbox{Im}\,
a_{\omega N\rightarrow\omega N}^{(\pi N)}\simeq 1.1$ fm, which
correspond to $\Delta\Gamma_\rho^{\,(\pi N)} = 65$ MeV and
$\Delta\Gamma_\omega^{\,(\pi N)} = 200$ MeV on the free mass shell (at
$\rho_N = \rho_0$). In this model, the $\pi N$ channel is responsible
for the major part of the in-medium width of the $\omega$ meson at
masses below the free one, while for the $\rho$ meson the $\pi\Delta$
channel dominates. As I discuss below, the $\pi N$ contributions can
be related to the experimentally accessible reactions $\pi N
\rightarrow V N$. Consequently, such reactions provide constraints on
models for the vector-meson self energy in nuclear matter at low
baryon densities.

\begin{figure}[t]
\center{\epsfig{file=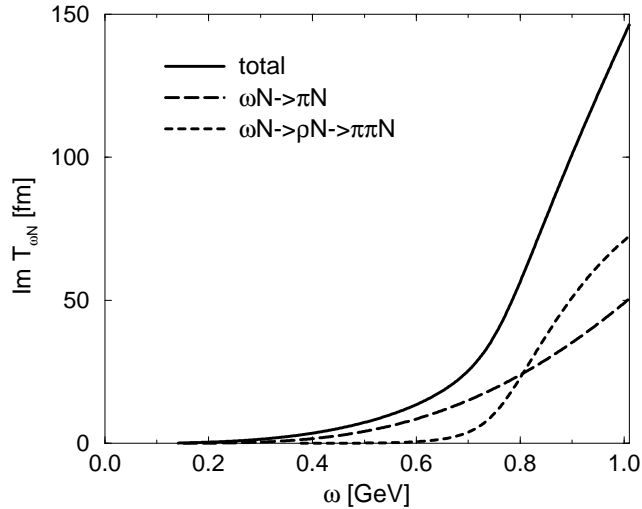,height=75mm}}
\caption{\label{Iome} Same as Fig.~\ref{Irho} for the
$\omega$-nucleon T-matrix.}
\end{figure}

The fact that the QCD sum rules are satisfied by the in-medium
$\rho$-meson spectral function of Hatsuda and
Lee~\cite{Hatsuda-Lee} and by the one of Klingl {\em et
al.}~\cite{KKW}, indicates that the $\rho$-meson width plays an
important role in this context. Indeed, Leupold {\em et
al.}~\cite{LPM} find that there is a trade-off between the
in-medium $\rho$-meson energy and its in-medium width. For a small
$\rho$ width, the QCD sum rule requires an appreciable downward
shift of the $\rho$ energy, while for a large width the sum rule is
satisfied with a small energy shift. Note that these results are
obtained with the vacuum saturation assumption for the 4-quark
condensate. Thus, the uncertainty connected with this approximation
remains~\cite{Hatsuda-Lee,KKW,LPM}.

\section{$\rho N$ interactions at finite $\vec{q}$}

So far I have discussed only vector mesons at rest in nuclear
matter. In a heavy-ion collision, the leptonic decay of such a
vector meson gives rise to a lepton pair of vanishing total
momentum in the rest frame of the participants. However,
so far the experimental invariant mass distributions are integrated
over the total momentum of the lepton pair. Consequently, in a
theoretical analysis of the lepton-pair spectrum in nucleus-nucleus
collisions the properties of vector mesons moving with respect to
the medium must be considered. In a recent paper H.J.~Pirner and
I~\cite{FP} estimated the interactions of $\rho$ mesons with
nucleons at finite momenta and  explored the consequences for the
production of lepton pairs in relativistic nucleus-nucleus
collisions.

As discussed in section 4 the $\rho$-meson self energy in
low-density nuclear matter can be constructed from the
$\rho$-nucleon scattering amplitude $f_{\rho N}$. In this work we
approximate the scattering amplitude with baryon resonance
contributions. There are two resonances in the mass range of
interest ($m \approx m_N + m_\rho$) which  couple strongly to the
$\rho N$ channel \cite{PDG}, namely $N^\star(1720)$ and $\Delta
(1905)$. The $N^\star(1720)$, which is the more important one,
decays into a $\rho$ meson and a nucleon in a relative p-wave with
more than 70~\% probability, while the $\Delta (1905)$ decays into
the $\rho N$ channel with a branching ratio above 60~\% in a
relative p- or f-wave. Thus, the corresponding $\rho$ self energy is
momentum dependent.

In order to describe the coupling of a virtual photon to hadrons we
adopt the formulation of the vector meson dominance (VMD) model due
to Kroll, Lee and Zumino \cite{KLZ}. This formulation of VMD has
the advantage of being explicitly gauge invariant, and it allows us
to adjust the $\gamma N$ and $\rho N$ partial widths independently.

The free parameters of the model are the $N^\star N\gamma$ and
$N^\star N\rho$ transition matrix elements together with the
corresponding ones for the $\Delta (1905)$. These are fixed by
fitting the partial widths for the resonance decays into the
$\gamma N$ and $\rho N$ channels. In the static
approximation~\cite{HFN}, the corresponding $\rho$-meson self
energy in nuclear matter is, to lowest order in density, given
by~\cite{FP}
\begin{eqnarray}
\label{static}
\Sigma_\rho(\omega,\vec{q}) - \Sigma^{(0)}_\rho(\omega,\vec{q})&=&
{4\over 3}{f_{N^\star N\rho}^2\over m_\rho^2} F(\vec{q}^{\,\, 2})
\vec{q}^{\,\, 2}\rho_B{(\varepsilon_q^{N^\star} - m_N) \over \omega^2 -
(\varepsilon_q^{N^\star} - m_N)^2}\\ &+& (N^\star \rightarrow
\Delta),\nonumber
\end{eqnarray}
where
\be
\label{energy}
\varepsilon_q^{N^\star} = \sqrt{\vec{q}^{\,\, 2} + m_{N^\star}} - {i\over
2}\Gamma_{N^\star},
\ee
$F(\vec{q}^{\,\, 2}) = \Lambda^2/(\Lambda^2 + \vec{q}^{\,\, 2})$
is a form factor with
$\Lambda = 1.5$ GeV and $\Sigma^{(0)}_\rho(\omega,\vec{q})$ denotes the
$\rho$-meson self energy in vacuum. In Eq.~(\ref{energy})
$\Gamma_{N^\star}$ is the full width of the resonance, modified by
the phase space appropriate for a resonance embedded in the
$\rho$-meson self energy.  This is important for the correct threshold
behaviour of the in-medium self energy. Note that in the static
approximation the self energy (\ref{static}) vanishes for a $\rho$
meson at rest in nuclear matter.

\begin{figure}[t]
\center{\epsfig{file=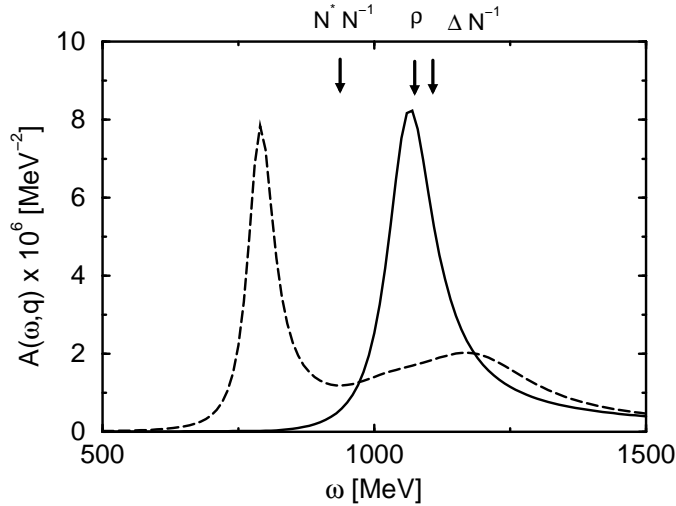,height=90mm,angle=-90}}
\caption{\label{fig:spectral}The $\rho$-meson spectral function
at $\mid\vec{q}\mid = 750$ MeV in vacuum (full line) and in nuclear matter at
$\rho_B = 2 \rho_0$ (from ref.~14). The arrows show the positions of
the unperturbed levels in the zero width limit.}
\end{figure}

The resulting spectral function is shown in Fig.~\ref{fig:spectral}
at $\mid\vec{q}\mid = 750$ MeV and $\rho_B = 2\rho_0$ together with the spectral
function of a $\rho$ meson in vacuum. Note that at finite momenta
strength is moved down to energies below the $\rho$-meson peak in
vacuum. Due to the energy dependence of the self energy and the
large widths of all particles involved, the spectral strength is
quite fragmented at large momenta. The peaked structure at low
energies, which one can identify with a quasiparticle, carries
only about 20 \% of the strength at $q = 750$ MeV. Nevertheless, it
may have interesting consequences for the lepton-pair spectrum in
nucleus-nucleus collisions.

\begin{figure}[t]
\center{\epsfig{file=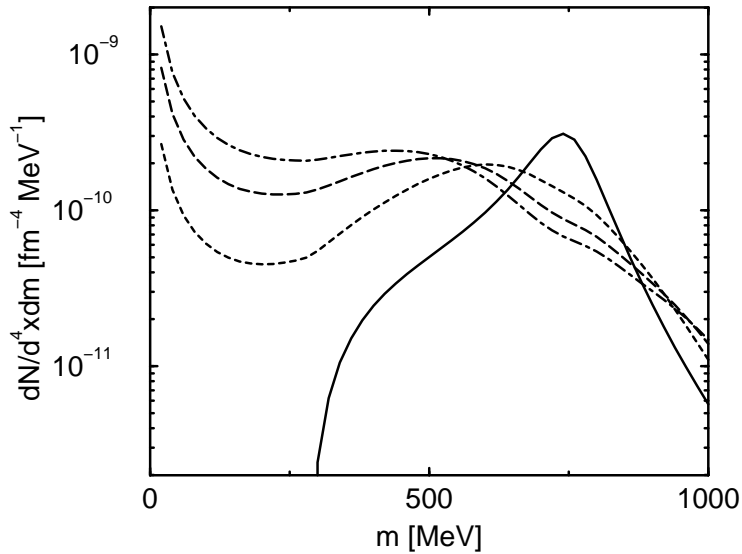,height=100mm,angle=-90}}
\caption{\label{fig:rate} The rate for lepton pair production
at $T = 140$ MeV and $\rho_B =$ 0 (full line), $\rho_0$ (dashed
line), $2 \rho_0$ (long-dashed line) and $3 \rho_0$ (dash-dotted
line) (from ref.~14).}
\end{figure}

In Fig.~\ref{fig:rate} the resulting production rate for $e^+e^-$
pairs is shown for various densities, at a temperature of $T = 140$
MeV. Also shown is the rate due to $\pi^+\pi^-$ annihilation computed
without any medium modifications. The rate includes an integral over
all pair momenta, but no corrections for experimental
acceptance. The p-wave interaction leads to a strong
enhancement of the population of lepton pairs with invariant masses
around 300-500 MeV. In the mass region of interest, the pion
annihilation and the $N^\star (1720)$ contributions dominate. Because
the p-wave interactions lead to a characteristic momentum dependence
of the lepton-pair spectrum, an analysis of the data in terms of total
momentum and invariant mass may shed light on the mechanism
responsible for the low-mass enhancement.

This calculation was recently extended by Peters {\em et
al.}~\cite{PLPM}, who include all baryon resonances of a mass below
1.9 GeV and attempt a self consistent treatment of the $\rho$-meson
spectral function. In the resulting spectral function, the
$\rho$-meson quasi-particle has acquired a very large width.
Similar results have been obtained by Rapp {\em et
al.}~\cite{CRW,CBRW}, who also compute the corresponding
lepton-pair spectrum for $S+Au$ and $Pb+Au$ collisions at CERN
energies, including the broadening of the $\rho$ meson in a nuclear
medium discussed in the previous section and the p-wave
interactions discussed above. This model provides a quantitative
description of the data of the CERES collaboration.

\section{Constraints from elementary processes}

The low-density theorem states that the self energy of e.g. a
vector meson $V$ in nuclear matter is given by~\cite{Lenz,DHL}
\begin{equation}
\Sigma_V(\rho_N) = -4\pi(1+\frac{m_V}{m_N})\langle{f}_{V N}\rangle\,\, \rho_N
+ \dots,
\label{LDT}
\end{equation}
where $m_V$ is the mass of the vector meson, $m_N$ that of the
nucleon, $\rho_N$ the nucleon density and $\langle{f}_{V N}\rangle$
denotes the $V N$ forward scattering amplitude $f_{V N}$,
appropriately averaged over the nucleon Fermi sea. Note that when
there is a narrow resonance close to threshold, the validity of the
low-density theorem may be limited to very low densities (see e.g.
ref.~\cite{Lutz}). For the vector mesons $\rho, \omega$ and $\phi$
the elastic scattering amplitudes have to be extracted indirectly,
from e.g. vector-meson production experiments, like $\gamma N
\rightarrow V N$ and $\pi N \rightarrow V N$.

In ref.~\cite{FS} a simple model for the photo-induced production of
$\rho$ and $\omega$ mesons was studied. Using the low density theorem
and vector meson dominance to extrapolate the amplitude from the
photon point $(q^2=0)$ to an on-shell $\rho$ meson $(q^2 = m_\rho^2)$,
one finds that the data are consistent with a strongly attractive
$\rho$-meson self energy in nuclear matter. However, the
extrapolation over a wide range in mass introduces a strong model
dependence. Although it may be possible to eliminate this model
dependence to some extent~\cite{zahed}, the safest approach is clearly
to constrain the $V N$ scattering amplitude only with data in the
relevant kinematic range. As an example, I shall discuss the
implications of the data on pion-induced vector-meson production for
the in-medium width of $\rho$ and $\omega$ mesons.

The spin-averaged cross section for the reaction $\pi^- p
\rightarrow \omega n$ is given by
\begin{equation}
\sigma_{\pi^- p \rightarrow \omega n} =
\frac{1}{2}\sum_{spins}\int\mid f_{\pi^- p \rightarrow \omega n}
\mid^2 \frac{k_\omega}{k_\pi}\mbox{d}\, \Omega,
\label{cross-sec}
\end{equation}
where $k_\omega$ and $k_\pi$ are the c.m. momenta in the $\omega N$
and $\pi N$ channels. Detailed balance implies a relation between
cross sections for time reversed reactions~\cite{LLQM}, which for the
case at hand means that
\begin{equation}
3 k_\omega^2 \sigma_{\omega n \rightarrow \pi^- p}
= k_\pi^2 \sigma_{\pi^- p \rightarrow \omega n}.
\label{detbal}
\end{equation}
The factor $3$ is the ratio of the spin degeneracies in the $\omega
N$ and $\pi N$ channels. Furthermore, it follows from (two-body)
unitarity that
\begin{equation}
\mbox{Im}\,f_{\omega n\rightarrow \omega n}(\theta=0)
= \sum _\nu k_\nu\int \mid f_{\omega n\rightarrow \nu}\mid^2\,
\frac{\mbox{d}\Omega}{4 \pi},
\label{unitarity}
\end{equation}
where $\mbox{Im}\,f_{\omega n\rightarrow \omega n}(\theta=0)$ is the
imaginary part of the $\omega$-nucleon forward scattering amplitude
and the sum runs over all possible two-body final states $\mid
\nu\rangle$. The contribution of the $\pi^- p$ channel is given by
\begin{equation}
\frac{4 \pi}{k_\pi}\mbox{Im}\,f_{\omega n\rightarrow
\omega n}^{\,(\pi^- p)}(\theta=0)
= \sum_{spins}\int\mid f_{\omega n\rightarrow \pi^- p}\mid^2
\mbox{d}\,\Omega,
\label{unitarity2}
\end{equation}
where the sum over spins applies only to the final state $\mid
\pi^- p\rangle$. Thus, using unitarity and detailed balance, one
finds
\begin{figure}[t]
\center{\epsfig{file=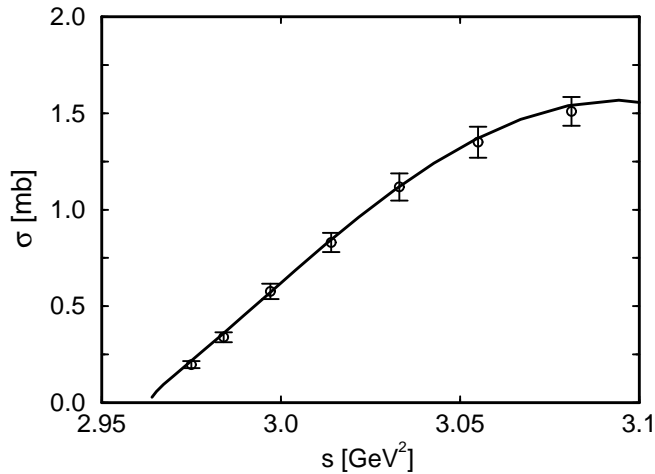,height=80mm}}
\caption{\label{omefit} The fit to the $\pi^- p
\rightarrow \omega n$ data (ref.~24) near threshold.}
\end{figure}
\begin{equation}
\sigma_{\pi^- p \rightarrow \omega n} =
12 \pi \frac{k_\omega}{k_\pi^2}\, \mbox{Im}\,\bar{f}_{\omega
n\rightarrow\omega n}^{\,(\pi^- p)}(\theta=0),
\label{pi-ome}
\end{equation}
where $\bar{f}$ denotes the spin-averaged scattering amplitude.
Close to the $\omega n$ threshold, the scattering amplitude can be
expanded in powers of the relative momentum in the $\omega n$
channel $k_\omega$. An excellent fit to the data from threshold up
to $s=3.1$ GeV$^2$ is obtained with $\mbox{Im}\,\bar{f}_{\omega
n\rightarrow\omega n}^{\,(\pi^- p)}=a + b k_\omega^2 + c
k_\omega^4$, where $a = 0.013$ fm, $b = 0.10$ fm$^3$ and $c = -
0.08$ fm$^5$ (see Fig.~\ref{omefit}). The coefficient $a$ is the
imaginary part of the scattering length, while $b$ may be due to a
p-wave contribution or a generalized effective range
term~\footnote{In principle there is a term linear in $k_\omega$,
whose coefficient is given in terms of the scattering
length~\cite{LLQM}: $-2 a^2$. However, this term is small and
consequently unimportant.}. Note that if the $k$-dependent terms are dropped,
i.e., if only the scattering length is kept, the energy dependence
of the cross section is too weak, and one cannot describe the data
at energies beyond the first two data points.

The contribution of the $\pi$-nucleon channel to the width of the
$\omega$ meson at rest in nuclear matter can now be obtained by
using the low-density theorem~(\ref{LDT})
\begin{equation}
\Delta\Gamma_\omega = 4\pi(1+\frac{m_\omega}{m_N})\frac{3}{2}
\frac{\langle \mbox{Im}\,
f_{\omega n\rightarrow\omega n}^{\,(\pi^- p)}\rangle
\,\rho_N}{m_\omega}.
\label{gam-ome}
\end{equation}
The factor $\frac{3}{2}$ accounts for the $\pi^0 n$ channel. At
nuclear matter density ($\rho_N = 0.16$ fm$^{-3}$) the average over
the Fermi sea yields an effective scattering amplitude $\langle
\mbox{Im}\,f_{\omega n\rightarrow\omega n}^{\,(\pi^- p)}\rangle =
0.031$ fm. This implies that the width of the $\omega$ meson in
nuclear matter is increased by 9 MeV due to the $\pi$-nucleon
channel~\footnote{Note that the imaginary part of the $\omega N$
scattering amplitude extracted from the data is surprisingly small.
The corresponding amplitude for $\eta N$ scattering, extracted from
the $\pi^- p\rightarrow \eta n$ data, is about an order of
magnitude larger: $\mbox{Im}\,a_{\eta n\rightarrow \eta n}^{(\pi^-
p)}= $ 0.2\,-\,0.3 fm (see e.g.~ref.~\cite{SFN,GW}).}.

A comparison of the imaginary part of the empirical scattering length
($1.5\times 0.02$ fm $=0.03$ fm) or the effective scattering amplitude
($1.5\times 0.034$ fm $=0.05$ with that extracted from the calculation
of Klingl {\em et al.}~\cite{KKW} in section 2 (1.1 fm) reveals a very
large discrepancy. In this model the imaginary part of the $\omega N$
scattering length and the in-medium width of the $\omega$ meson are
overestimated by more than an order of magnitude! This is a striking
illustration of how dangerous an extrapolation over a wide range in
energy can be, without guidance from experiment.
\begin{figure}[t]
\center{\epsfig{file=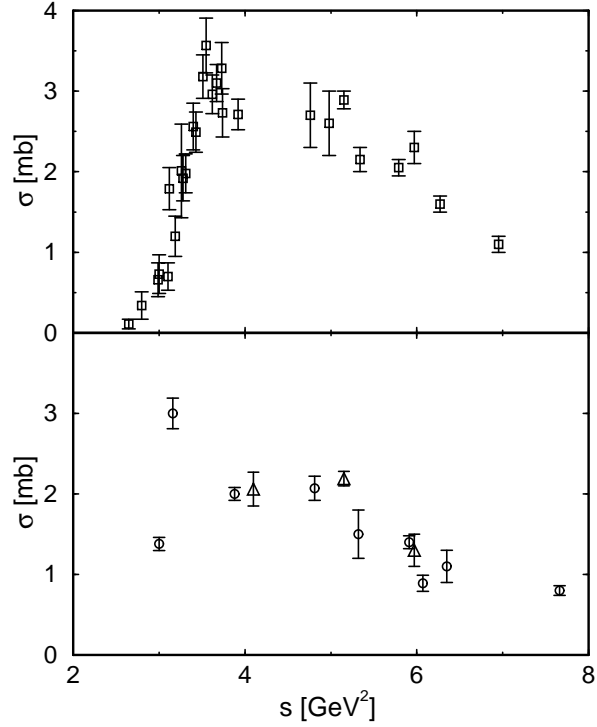,height=100mm}}
\caption{\label{rhodata} The data for the reactions
$\pi^- p \rightarrow \rho^0 n$(squares), $\pi^+ p
\rightarrow \rho^+ p$ (circles) and $\pi^- p \rightarrow
\rho^- p$ (triangles) (ref.~24).}
\end{figure}

For the $\rho$ meson the situation is a bit more complicated. First
of all the experimentally accessible $\pi N$ channel is
subdominant. Second, both isospin 1/2 and 3/2 are allowed. Thus,
three independent reactions are needed to pin down the amplitudes
of the two isospin channels and their relative phase. For the
$\omega$ meson the situation is simpler, since only isospin 1/2 is
allowed. The available data~\cite{lanbor} on the reactions $\pi^-
p\rightarrow\rho^0 n$, $\pi^+ p\rightarrow \rho^+ p$ and $\pi^-
p\rightarrow \rho^- p$ up to $s = 4$ GeV$^2$ are shown in
Fig.~\ref{rhodata}. These cross sections involve linearly
independent combinations of the isospin amplitudes and would, if
measured down to threshold, be sufficient to determine the
amplitudes. The enhancement of the $\rho$-meson width in nuclear
matter is then given by
\begin{eqnarray}
\label{gamma-rho}
\Delta\Gamma_\rho &=& \frac{4\pi}{2 m_\rho}(1+\frac{m_\rho}{m_N})
\left(\langle \mbox{Im}\,f_{\rho^0 n\rightarrow\rho^0 n}^{\,(\pi^-
p)}\rangle + \langle \mbox{Im}\,f_{\rho^+ p\rightarrow\rho^+
p}^{\,(\pi^+ p)}\rangle\right.\\\nonumber &+& \left. \langle
\mbox{Im}\,f_{\rho^- p\rightarrow\rho^- p}^{\,(\pi^- p)}\rangle\right)
\,\rho_N.
\end{eqnarray}
Unfortunately only the first one, $\pi^- p\rightarrow \rho^0 n$, is
well known close to threshold. For the second one, $\pi^+ p\rightarrow
\rho^+ p$, only two points are available in the relevant energy range,
while the data on $\pi^- p\rightarrow \rho^- p$ start only at $s >
4$ GeV$^2$. Thus, in order to extract numbers, we must make
assumptions on the cross section for the last reaction close to
threshold. Clearly data on $\pi^+ p\rightarrow \rho^+ p$ and $\pi^-
p\rightarrow \rho^- p$ close to threshold would be very useful.

\begin{figure}[t]
\center{\epsfig{file=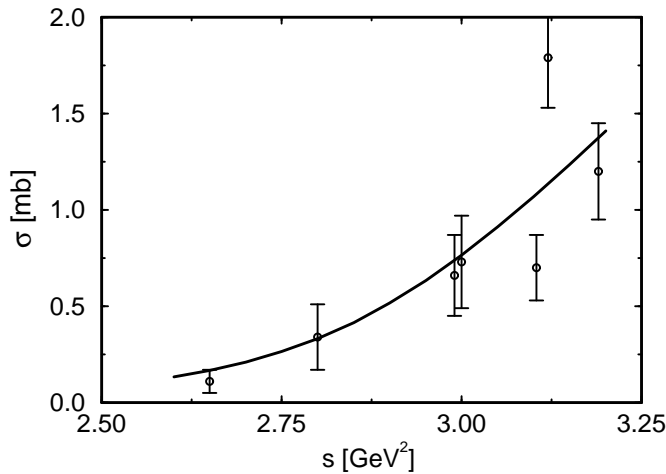,height=80mm}}
\caption{\label{rhofit} The data for the reaction
$\pi^- p \rightarrow \rho^0 n$ near threshold and the fit described
in the text.}
\end{figure}

Because the $\rho$-meson width is large, the detailed-balance relation
is more complicated ref.~\cite{BerDan}. One must take into account
the fact that at a given c.m. energy $\sqrt{s}$ part of the $\rho$-meson
spectral strength may be energetically unavailable. Thus, the cross
section for production of $\rho$ mesons is of the form
\begin{equation}
\label{pi-rho}
\sigma_{\pi^- p \rightarrow \rho^0 n} =
12 \pi \int_{2 m_\pi}^{\sqrt{s} - m_N}\frac{k_\rho}{k_\pi^2}\,
\mbox{Im}\,\bar{f}_{\rho^0 n\rightarrow\rho^0 n}^{\,(\pi^- p)}\, A_\rho(m^2)
\frac{\mbox{d}\,m^2}{\pi}.
\end{equation}
Here $A_\rho(m^2)$ is the $\rho$-meson spectral function, and
$k_\rho^2=((s-m^2-m_N^2)^2-4m_N^2m^2)/4s$. For a narrow resonance,
where the spectral function can be approximated by a delta function
$A_R(m^2)\simeq \delta(m^2-m_R^2)$, Eq.~(\ref{pi-rho}) reduces to
the standard form (\ref{pi-ome}). A good fit to the $\pi^- p
\rightarrow \rho^0 n$ data near threshold is obtained with
$\mbox{Im}\,\bar{f}_{\rho^0 n\rightarrow\rho^0 n}^{\,(\pi^- p)}
= a + b k_\rho^2$, where $a=0.021$ fm and $b=0.006$ fm$^{3}$
(see Fig.~\ref{rhofit}).
By averaging over the Fermi sea, I find an effective scattering
amplitude $\langle\mbox{Im}\bar{f}_{\rho^0 n\rightarrow\rho^0
n}^{\,(\pi^- p)}\rangle = 0.022$ fm.

Similarly, a fit to the low-energy $\pi^+ p\rightarrow \rho^+ p$
data can be obtained. Because there are only two data points in the
energy range of interest, it makes no sense to extract a momentum
dependent scattering amplitude. A reasonable fit is obtained with
$\mbox{Im}\,\bar{f}_{\rho^+ p\rightarrow \rho^+ p}^{\,(\pi^+ p)} =
0.06-0.08$ fm.

One can use the isospin decomposition of the $\pi N\rightarrow \rho
N$ scattering amplitude and the two measured cross sections to set
limits on the magnitude of the unknown $\pi^- p \rightarrow\rho^-
p$ amplitude~\footnote{In this argument I neglect the momentum
dependence of the amplitudes, which for the reaction
$\pi^-p\rightarrow \rho^0 n$ gives rise to at most a 10 \%
correction at the momenta of interest here.}. The absolute value of
the isospin 3/2 $\pi N\rightarrow\rho N$ amplitude is fixed by the
$\pi^+ p\rightarrow \rho^+ p$ data, while the $\pi^- p\rightarrow
\rho^0 n$ data give a relation between the isospin 1/2 amplitude
and the relative phase of the amplitudes. This relation implies
limits on the magnitude of the isospin 1/2 amplitude, and
consequently also constraints on the $\pi^-p\rightarrow \rho^- p$
amplitude. One thus finds that the absolute value of the isospin
1/2 amplitude must be less than 0.35 fm, which implies for the
$\langle\mbox{Im}\,f_{\rho^- p\rightarrow\rho^- p}^{\,(\pi^-
p)}\rangle < 0.25$ fm and that the effective ``scattering length''
due to the $\pi N$ channel fulfills
\begin{equation}
\label{esl}
\mbox{Im}\,a_{eff}^{(\pi N)}=\frac{\langle
\mbox{Im}\,f_{\rho^0 n\rightarrow\rho^0 n}^{\,(\pi^- p)}\rangle +
\langle \mbox{Im}\,f_{\rho^+ p\rightarrow\rho^+ p}^{\,(\pi^+ p)}\rangle
+ \langle \mbox{Im}\,f_{\rho^- p\rightarrow\rho^- p}^{\,(\pi^-
p)}\rangle}{2} < 0.18\, \mbox{fm}.
\end{equation}

This constraint is not very restrictive, since it allows the
unknown scattering amplitude near threshold to be an order of
magnitude larger than that for the reaction $\pi^-
p\rightarrow\rho^0 n$. To proceed further one must make an
assumption for $\langle\mbox{Im}\,f_{\rho^- p\rightarrow\rho^-
p}^{\,(\pi^- p)}\rangle$. At energies, where all reaction channels
are measured ($s > 4$ GeV$^2$), the cross sections agree within a
factor $\sim 2$. Hence, a reasonable guess is that this amplitude
is not very different from the ones extracted for the other
reactions. If one assumes that the unknown scattering amplitude is
between those obtained for the reactions $\pi^- p\rightarrow \rho^0
n$ and $\pi^+ p\rightarrow \rho^+ p$, one finds that the $\pi N$
contribution to $\mbox{Im}\,a_{eff}^{(\pi N)} = 0.05-0.09$ fm. This
corresponds to an enhancement of the $\rho$-meson width of
$\Delta\Gamma_\rho^{(\pi N)} = 9-17$ MeV at nuclear matter density.
A comparison with the value extracted for the model of Klingl {\em
et al.} (0.35 fm), reveals a discrepancy of a factor $4-7$. In
order to reproduce their value, the unknown scattering amplitude
would have to be $\langle\mbox{Im}\,f_{\rho^- p\rightarrow\rho^-
p}^{\,(\pi^- p)}\rangle=0.61$~fm, which clearly violates the
upper limit derived above.

In ref.~\cite{KKW} it is found that the imaginary part of the $\rho
N$ T-matrix is dominated by the experimentally not accessible
$\Delta\pi$ channel. If this is the case, one cannot draw any firm
conclusions on the in-medium width of the $\rho$ meson.
Nevertheless, one can make qualitative statements. One does not
expect the matrix element for $\rho N\rightarrow
\pi\Delta$ to differ considerably from that for $\rho
N\rightarrow \pi N$, since the $\pi\Delta N$ coupling constant is
believed to be about a factor 2 larger than the $\pi N N$ one (cf.
ref.~\cite{BW}).  Furthermore, the spin and isospin factors favour
the $\Delta\pi$ channel, while the phase space factor obviously is
larger in the $\pi N$ channel. To the extent that the latter two
effects cancel, one would expect a net factor of approximately 4.
This agrees approximately with the relative strength of the two
channels found by Klingl {\em et al.} for an on-shell $\rho$ meson
(see Fig.~\ref{Irho}). Consequently, if we assume that this ratio
is correct, one finds for the total scattering length
$\mbox{Im}\,a_{eff}^{(tot)} \simeq 5\,\mbox{Im}\,a_{eff}^{(\pi N)}
= 0.25-0.45$~fm and for the enhancement of the $\rho$-meson width,
including both channels, $\Delta\Gamma_\rho = 45-85$ MeV at $\rho_N
= \rho_0$, while Klingl {\em et al.} find 1.62 fm, which implies
$\Delta\Gamma_\rho = 300$ MeV.

Note that if the estimate given above is correct and the total
width of the $\rho$ meson at nuclear matter density is $\simeq 220$
MeV (on the free mass shell), the QCD sum rule requires a
strong reduction of the $\rho$-meson energy at this
density~\cite{LPM} (modulo uncertainties due to the approximate
treatment of the 4-quark condensate).

Also in the work of Herrmann {\em et al.}~\cite{HFN}, where the
broadening of the $\rho$ meson in nuclear matter is due to the $\Delta
\pi$ channel, the enhancement of the width, $\Delta\Gamma_\rho\simeq
150$ MeV at nuclear matter density, is appreciably larger than the
estimate given above. A similar model is used by Rapp {\em et
al.}~\cite{CRW,CBRW} in their calculations of lepton-pair
production in nucleus-nucleus collisions. Part of the low-mass
enhancement is in this model due to the broadening of the $\rho$
meson in nuclear matter. Clearly any serious attempt to understand
the low mass lepton pairs should be confronted with the vector
meson production data discussed here.

\section{Summary}

I have discussed recent theoretical developments on the many-body
effects, which determine the properties of vector mesons in nuclear
matter and the consequences for the lepton pair spectrum in
hadronic collisions. The broadening of $\rho$ mesons in nuclear
matter and the momentum dependence of the $\rho$-meson self energy
leads to an enhancement of low-mass lepton pairs. In the work of
Rapp {\em et al.}~\cite{CRW,CBRW} these effects are responsible for
most of the low-mass enhancement found by the CERES~\cite{CERES}
and Helios-3~\cite{Helios} collaborations at CERN. Thus, the
many-body processes offer an interpretation of the low-mass
enhancement alternative to the dropping-mass scenario of Li, Ko and
Brown~\cite{likobr}.

However, elementary processes put constraints on the in-medium
width of $\rho$ and $\omega$ mesons, which so far have not been
implemented in the calculations of vector-meson properties in
nuclear matter. The widths extracted from the data on pion-induced
vector-meson production are much smaller than those obtained within
various models. Since, in the many-body scenario, a substantial
part of the low-mass enhancement is due to the in-medium broadening
of the $\rho$ meson, this is a crucial point. How these constraints
modify the lepton-pair spectrum in this approach is presently being
investigated~\cite{RW}.

I argued that the root of the problem in some models is the
potentially dangerous extrapolation from the photon point to
$q^2=m_V^2$. Consequently, a safer approach would be to exclude the
photon data and to rely only on hadron scattering and production
data in the relevant kinematic regime.

Finally, I note that Klingl {\em et al.} have now revised their calculation,
so that the $\pi N$ channel is in agreement with the
$\omega$-production data~\cite{KKW2}. In the revised model they also
include processes that were not included in their original
calculation. These processes enhance the contribution of the $\pi\pi
N$ channel to the in-medium width of the $\omega$ meson. Thus, Klingl
{\em at al.} now find that, at $\rho = \rho_0$, the $\omega$ width is
enhanced by $\simeq 30$ MeV and its mass is reduced by $\simeq 120$
MeV.

\section*{Acknowledgments}

I am grateful to F.~Klingl, M.~Lutz, W.~Weise and G.~Wolf for
valuable discussions.

\end{document}